\documentclass[12pt]{article}
\usepackage{amssymb}
\usepackage{amsmath}

\begin{document}

\begin{titlepage}
\title{\bf Hamiltonian Mechanic Systems on the Standard Cliffordian K\"{a}hler Manifolds}
\author{ Mehmet Tekkoyun \footnote{Corresponding author. E-mail address: tekkoyun@pau.edu.tr; Tel: +902582953616; Fax: +902582953593}\\
{\small Department of Mathematics, Pamukkale University,}\\
{\small 20070 Denizli, Turkey}}
\date{\today}
\maketitle

\begin{abstract}

This study introduces standard Cliffordian  K\"{a}hler analogue of
Hamiltonian mechanic systems. In the end, the some results related
to standard Cliffordian K\"{a}hler dynamical systems are also
discussed.

{\bf Keywords:} Cliffordian K\"{a}hler Geometry, Hamiltonian
Mechanic Systems.

{\bf PACS:} 02.40.

\end{abstract}
\end{titlepage}

\section{Introduction}

It is well-known that modern differential geometry explains explicitly the
dynamics of Hamiltonians. Therefore, if $Q$ is an $m$-dimensional
configuration manifold and $\mathbf{H}:T^{\ast }Q\rightarrow \mathbf{R}$%
\textbf{\ }is a regular Hamiltonian function, then there is a unique vector
field $X$ on $T^{\ast }Q$ such that dynamic equations are given by
\begin{equation}
\,\,i_{X}\Phi =d\mathbf{H}  \label{1.1}
\end{equation}%
where $\Phi $ indicates the symplectic form. The triple $(T^{\ast }Q,\Phi
,X) $ are called \textit{Hamiltonian system }on the cotangent bundle $%
T^{\ast }Q. $

Nowadays, there are a lot of studies about Hamiltonian mechanics,
formalisms, systems and equations \cite{deleon, tekkoyun} and there in.
There are real, complex, paracomplex and other analogues. We say that in
order to obtain different analogous in different spaces is possible.

Quaternions were invented by Sir William Rowan Hamiltonian as an extension
to the complex numbers. Hamiltonian's defining relation is most succinctly
written as:

\begin{equation}
i^{2}=j^{2}=k^{2}=ijk=-1  \label{1.2}
\end{equation}%
If it is compared to the calculus of vectors, quaternions have slipped into
the realm of obscurity. They do however still find use in the computation of
rotations. A lot of physical laws in classical, relativistic, and quantum
mechanics can be written pleasantly by means of quaternions. Some physicists
hope they will find deeper understanding of the universe by restating basic
principles in terms of quaternion algebra. It is well-known that quaternions
are useful for representing rotations in both quantum and classical
mechanics \cite{dan} . We say that Cliffordian manifold is quaternion
manifold. Therefore, all properties defined on quaternion manifold of
dimension $8n$ also is valid for Cliffordian manifold. Thus, it is possible
to construct mechanical equations on Cliffordian K\"{a}hler manifold.

The paper is structured as follows. In second 2, we recall Cliffordian K\"{a}%
hler manifolds. In second 3 we introduce Hamiltonian equations related to
mechanical systems on Cliffordian K\"{a}hler manifold. In conclusion, we
discuss some geometrical and physical results about Hamiltonian equations
and fields obtained on the base manifold.

\section{Preliminaries}

Hereafter, all mappings and manifolds are assumed to be smooth, i.e.
infinitely differentiable and the sum is taken over repeated indices. By $%
\mathcal{F}(M)$, $\chi (M)$ and $\Lambda ^{1}(M)$ we denote the set of
functions on $M$, the set of vector fields on $M$ and the set of 1-forms on $%
M$, respectively.

\subsection{Cliffordian K\"{a}hler Manifolds}

Here, we recalled the main concepts and structures given in \cite{yano,
burdujan} . Let $M$ be a real smooth manifold of dimension $m.$ Suppose that
there is a 6-dimensional vector bundle $V$ consisting of $F_{i}(i=1,2,...,6)$
tensors of type (1,1) over $M.$ Such a local basis $\{F_{1},F_{2},...,F_{6}%
\} $ is called a canonical local basis of the bundle $V$ in a neighborhood $%
U $ of $M$. Then $V$ is called an almost Cliffordian structure in $M$. The
pair $(M,V)$ is named an almost Cliffordian manifold with $V$. Hence, an
almost Cliffordian manifold $M$ is of dimension $m=8n.$ If there exists on $%
(M,V)$ a global basis $\{F_{1},F_{2},...,F_{6}\},$ then $(M,V)$ is said to
be an almost Cliffordian manifold; the basis $\{F_{1},F_{2},...,F_{6}\}$ is
called a global basis for $V$.

An almost Cliffordian connection on the almost Cliffordian manifold $(M,V)$
is a linear connection $\nabla $ on $M$ which preserves by parallel
transport the vector bundle $V$. This means that if $\Phi $ is a
cross-section (local-global) of the bundle $V$, then $\nabla _{X}\Phi $ is
also a cross-section (local-global, respectively) of $V$, $X$ being an
arbitrary vector field of $M$.

If for any canonical basis $\{J_{1},J_{2},...,J_{6}\}$ of $V$ in a
coordinate neighborhood $U$, the identities
\begin{equation}
g(J_{i}X,J_{i}Y)=g(X,Y),\text{ }\forall X,Y\in \chi (M),\text{ }\
i=1,2,...,6,  \label{2.2}
\end{equation}%
hold, the triple $(M,g,V)$ is named an almost Cliffordian Hermitian manifold
or metric Cliffordian manifold denoting by $V$ an almost Cliffordian
structure $V$ and by $g$ a Riemannian metric and by $(g,V)$ an almost
Cliffordian metric structure$.$

Since each $J_{i}(i=1,2,...,6)$ is almost Hermitian structure\ with respect
to $g$, setting

\begin{equation}
\Phi _{i}(X,Y)=g(J_{i}X,Y),~\text{ }i=1,2,...,6,  \label{2.3}
\end{equation}

for any vector fields $X$ and $Y$, we see that $\Phi _{i}$ are 6 local
2-forms.

If the Levi-Civita connection $\nabla =\nabla ^{g}$ on $(M,g,V)$ preserves
the vector bundle $V$ by parallel transport, then $(M,g,V)$ is called a
Cliffordian K\"{a}hler manifold, and an almost Cliffordian structure $\Phi
_{i}$ of $M$ is called a Cliffordian K\"{a}hler structure. A Clifford K\"{a}%
hler manifold is Riemannian manifold ($M^{8n},g$)$.$ For example, we say
that $\mathbf{R}^{8n}$ is the simplest example of Clifford K\"{a}hler
manifold. Suppose that let $\left\{
x_{i},x_{n+i},x_{2n+i},x_{3n+i},x_{4n+i},x_{5n+i},x_{6n+i},x_{7n+i}\right\}
, $ $i=\overline{1,n}$ be a real coordinate system on $\mathbf{R}^{8n}.$
Then we define by $\left\{ \frac{\partial }{\partial x_{i}},\frac{\partial }{%
\partial x_{n+i}},\frac{\partial }{\partial x_{2n+i}},\frac{\partial }{%
\partial x_{3n+i}},\frac{\partial }{\partial x_{4n+i}},\frac{\partial }{%
\partial x_{5n+i}},\frac{\partial }{\partial x_{6n+i}},\frac{\partial }{%
\partial x_{7n+i}}\right\} $ and $%
\{dx_{i},dx_{n+i},dx_{2n+i},dx_{3n+i},dx_{4n+i},dx_{5n+i},dx_{6n+i},dx_{7n+i}\}
$ be natural bases over $\mathbf{R}$ of the tangent space $T(\mathbf{R}%
^{8n}) $ and the cotangent space $T^{\ast }(\mathbf{R}^{8n})$ of $\mathbf{R}%
^{8n},$ respectively$.$ By structures $J_{1},J_{2},J_{3}$, the following
expressions are obtained%
\begin{equation}
\begin{array}{c}
J_{1}(\frac{\partial }{\partial x_{i}})=\frac{\partial }{\partial x_{n+i}},%
\text{ }J_{1}(\frac{\partial }{\partial x_{n+i}})=-\frac{\partial }{\partial
x_{i}},\text{ }J_{1}(\frac{\partial }{\partial x_{2n+i}})=\frac{\partial }{%
\partial x_{4n+i}},\text{ }J_{1}(\frac{\partial }{\partial x_{3n+i}})=\frac{%
\partial }{\partial x_{5n+i}}, \\
J_{1}(\frac{\partial }{\partial x_{4n+i}})=-\frac{\partial }{\partial
x_{2n+i}},\text{ }J_{1}(\frac{\partial }{\partial x_{5n+i}})=-\frac{\partial
}{\partial x_{3n+i}},\text{ }J_{1}(\frac{\partial }{\partial x_{6n+i}})=%
\frac{\partial }{\partial x_{7n+i}},\text{ }J_{1}(\frac{\partial }{\partial
x_{7n+i}})=-\frac{\partial }{\partial x_{6n+i}}, \\
J_{2}(\frac{\partial }{\partial x_{i}})=\frac{\partial }{\partial x_{2n+i}},%
\text{ }J_{2}(\frac{\partial }{\partial x_{n+i}})=-\frac{\partial }{\partial
x_{4n+i}},\text{ }J_{2}(\frac{\partial }{\partial x_{2n+i}})=-\frac{\partial
}{\partial x_{i}},\text{ }J_{2}(\frac{\partial }{\partial x_{3n+i}})=\frac{%
\partial }{\partial x_{6n+i}}, \\
J_{2}(\frac{\partial }{\partial x_{4n+i}})=\frac{\partial }{\partial x_{n+i}}%
,\text{ }J_{2}(\frac{\partial }{\partial x_{5n+i}})=-\frac{\partial }{%
\partial x_{7n+i}},\text{ }J_{2}(\frac{\partial }{\partial x_{6n+i}})=-\frac{%
\partial }{\partial x_{3n+i}},\text{ }J_{2}(\frac{\partial }{\partial
x_{7n+i}})=\frac{\partial }{\partial x_{5n+i}}, \\
J_{3}(\frac{\partial }{\partial x_{i}})=\frac{\partial }{\partial x_{3n+i}},%
\text{ }J_{3}(\frac{\partial }{\partial x_{n+i}})=-\frac{\partial }{\partial
x_{5n+i}},\text{ }J_{3}(\frac{\partial }{\partial x_{2n+i}})=-\frac{\partial
}{\partial x_{6n+i}},\text{ }J_{3}(\frac{\partial }{\partial x_{3n+i}})=-%
\frac{\partial }{\partial x_{i}}, \\
J_{3}(\frac{\partial }{\partial x_{4n+i}})=\frac{\partial }{\partial x_{7n+i}%
},\text{ }J_{3}(\frac{\partial }{\partial x_{5n+i}})=\frac{\partial }{%
\partial x_{n+i}},\text{ }J_{3}(\frac{\partial }{\partial x_{6n+i}})=\frac{%
\partial }{\partial x_{2n+i}},\text{ }J_{3}(\frac{\partial }{\partial
x_{7n+i}})=-\frac{\partial }{\partial x_{4n+i}}.%
\end{array}
\label{2.4}
\end{equation}

A canonical local basis$\{J_{1}^{\ast },J_{2}^{\ast },J_{3}^{\ast }\}$ of $%
V^{\ast }$ of the cotangent space $T^{\ast }(M)$ of manifold $M$ satisfies
the condition as follows:

\begin{equation}
J_{1}^{\ast 2}=J_{2}^{\ast 2}=\text{ }J_{3}^{\ast 2}=J_{1}^{\ast
}J_{2}^{\ast }\text{ }J_{3}^{\ast 2}J_{2}^{\ast }J_{1}^{\ast }=-I,
\label{2.6}
\end{equation}%
defining by%
\begin{equation}
\begin{array}{c}
J_{1}^{\ast }(dx_{i})=dx_{n+i},\text{ }J_{1}^{\ast }(dx_{n+i})=-dx_{i},\text{
}J_{1}^{\ast }(dx_{2n+i})=dx_{4n+i},\text{ }J_{1}^{\ast
}(dx_{3n+i})=dx_{5n+i}, \\
J_{1}^{\ast }(dx_{4n+i})=-dx_{2n+i},\text{ }J_{1}^{\ast
}(dx_{5n+i})=-dx_{3n+i},\text{ }J_{1}^{\ast }(dx_{6n+i})=dx_{7n+i},\text{ }%
J_{1}^{\ast }(dx_{7n+i})=-dx_{6n+i} \\
J_{2}^{\ast }(dx_{i})=dx_{2n+i},\text{ }J_{2}^{\ast }(dx_{n+i})=-dx_{4n+i},%
\text{ }J_{2}^{\ast }(dx_{2n+i})=-dx_{i},\text{ }J_{2}^{\ast
}(dx_{3n+i})=dx_{6n+i}, \\
J_{2}^{\ast }(dx_{4n+i})=dx_{n+i},\text{ }J_{2}^{\ast
}(dx_{5n+i})=-dx_{7n+i},\text{ }J_{2}^{\ast }(dx_{6n+i})=-dx_{3n+i},\text{ }%
J_{2}^{\ast }(dx_{7n+i})=dx_{5n+i}, \\
J_{3}^{\ast }(dx_{i})=dx_{3n+i},\text{ }J_{3}^{\ast }(dx_{n+i})=-dx_{5n+i},%
\text{ }J_{3}^{\ast }(dx_{2n+i})=-dx_{6n+i},\text{ }J_{3}^{\ast
}(dx_{3n+i})=-dx_{i}, \\
J_{3}^{\ast }(dx_{4n+i})=dx_{7n+i},\text{ }J_{3}^{\ast }(dx_{5n+i})=dx_{n+i},%
\text{ }J_{3}^{\ast }(dx_{6n+i})=dx_{2n+i},\text{ }J_{3}^{\ast
}(dx_{7n+i})=-dx_{4n+i}.%
\end{array}
\label{2.7}
\end{equation}

\section{Hamiltonian Mechanics}

Here, we obtain Hamiltonian equations and Hamiltonian mechanical system for
quantum and classical mechanics structured on the standard Cliffordian K\"{a}%
hler manifold $(\mathbf{R}^{8n},V).$

Firstly, let $(\mathbf{R}^{8n},V)$ be a standard Cliffordian K\"{a}hler
manifold. Assume that a component of almost Cliffordian structure $V^{\ast }$%
, a Liouville form and a 1-form on the standard Cliffordian K\"{a}hler
manifold $(\mathbf{R}^{8n},V)$ are shown by $J_{1}^{\ast }$, $\lambda
_{J_{1}^{\ast }}$ and $\omega _{J_{1}^{\ast }}$, respectively$.$

Then
\begin{eqnarray*}
\omega _{J_{1}^{\ast }} &=&\frac{1}{2}%
(x_{i}dx_{i}+x_{n+i}dx_{n+i}+x_{2n+i}dx_{2n+i}+x_{3n+i}dx_{3n+i} \\
&&+x_{4n+i}dx_{4n+i}+x_{5n+i}dx_{5n+i}+x_{6n+i}dx_{6n+i}+x_{7n+i}dx_{7n+i})
\end{eqnarray*}%
and
\begin{eqnarray*}
\lambda _{J_{1}^{\ast }} &=&J_{1}^{\ast }(\omega _{J_{1}^{\ast }})=\frac{1}{2%
}(x_{i}dx_{n+i}-x_{n+i}dx_{i}+x_{2n+i}dx_{4n+i}+x_{3n+i}dx_{5n+i} \\
&&-x_{4n+i}dx_{2n+i}-x_{5n+i}dx_{3n+i}+x_{6n+i}dx_{7n+i}-x_{7n+i}dx_{6n+i}).
\end{eqnarray*}%
It is well-known that if $\Phi _{J_{1}^{\ast }}$ is a closed K\"{a}hler form
on the standard Cliffordian K\"{a}hler manifold $(\mathbf{R}^{8n},V),$ then $%
\Phi _{J_{1}^{\ast }}$ is also a symplectic structure on Cliffordian K\"{a}%
hler manifold $(\mathbf{R}^{8n},V)$.

Consider that Hamilton vector field $X$ associated with Hamiltonian energy $%
\mathbf{H}$ is given by%
\begin{equation}
\begin{array}{c}
X=X^{i}\frac{\partial }{\partial x_{i}}+X^{n+i}\frac{\partial }{\partial
x_{n+i}}+X^{2n+i}\frac{\partial }{\partial x_{2n+i}}+X^{3n+i}\frac{\partial
}{\partial x_{3n+i}} \\
+X^{4n+i}\frac{\partial }{\partial x_{4n+i}}+X^{5n+i}\frac{\partial }{%
\partial x_{5n+i}}+X^{6n+i}\frac{\partial }{\partial x_{6n+i}}+X^{7n+i}\frac{%
\partial }{\partial x_{7n+i}}.%
\end{array}
\label{4.2}
\end{equation}

Then
\begin{equation}
\Phi _{J_{1}^{\ast }}=-d\lambda _{J_{1}^{\ast }}=dx_{n+i}\wedge
dx_{i}+dx_{4n+i}\wedge dx_{2n+i}+dx_{5n+i}\wedge dx_{3n+i}+dx_{7n+i}\wedge
dx_{6n+i}  \label{4.3}
\end{equation}%
and%
\begin{equation}
\begin{array}{c}
i_{X}\Phi _{J_{1}^{\ast }}=\Phi _{J_{1}^{\ast
}}(X)=X^{n+i}dx_{i}-X^{i}dx_{n+i}+X^{4n+i}dx_{2n+i}-X^{2n+i}dx_{4n+i} \\
+X^{5n+i}dx_{3n+i}-X^{3n+i}dx_{5n+i}+X^{7n+i}dx_{6n+i}-X^{6n+i}dx_{7n+i}.%
\end{array}
\label{4.4}
\end{equation}
Moreover, the differential of Hamiltonian energy is obtained as follows:%
\begin{equation}
\begin{array}{c}
d\mathbf{H}=\frac{\partial \mathbf{H}}{\partial x_{i}}dx_{i}+\frac{\partial
\mathbf{H}}{\partial x_{n+i}}dx_{n+i}+\frac{\partial \mathbf{H}}{\partial
x_{2n+i}}dx_{2n+i}+\frac{\partial \mathbf{H}}{\partial x_{3n+i}}dx_{3n+i} \\
+\frac{\partial \mathbf{H}}{\partial x_{4n+i}}dx_{4n+i}+\frac{\partial
\mathbf{H}}{\partial x_{5n+i}}dx_{5n+i}+\frac{\partial \mathbf{H}}{\partial
x_{6n+i}}dx_{6n+i}+\frac{\partial \mathbf{H}}{\partial x_{7n+i}}dx_{7n+i}.%
\end{array}
\label{4.5}
\end{equation}
According to \textbf{Eq.}(\ref{1.1}), if equaled \textbf{Eq. }(\ref{4.4})
and \textbf{Eq. }(\ref{4.5}), the Hamiltonian vector field is found as
follows:%
\begin{equation}
\begin{array}{c}
X=-\frac{\partial \mathbf{H}}{\partial x_{n+i}}\frac{\partial }{\partial
x_{i}}+\frac{\partial \mathbf{H}}{\partial x_{i}}\frac{\partial }{\partial
x_{n+i}}-\frac{\partial \mathbf{H}}{\partial x_{4n+i}}\frac{\partial }{%
\partial x_{2n+i}}-\frac{\partial \mathbf{H}}{\partial x_{5n+i}}\frac{%
\partial }{\partial x_{3n+i}} \\
+\frac{\partial \mathbf{H}}{\partial x_{2n+i}}\frac{\partial }{\partial
x_{4n+i}}+\frac{\partial \mathbf{H}}{\partial x_{3n+i}}\frac{\partial }{%
\partial x_{5n+i}}-\frac{\partial \mathbf{H}}{\partial x_{7n+i}}\frac{%
\partial }{\partial x_{6n+i}}+\frac{\partial \mathbf{H}}{\partial x_{6n+i}}%
\frac{\partial }{\partial x_{7n+i}}%
\end{array}
\label{4.6}
\end{equation}

Suppose that a curve
\begin{equation}
\alpha :\mathbf{R}\rightarrow \mathbf{R}^{8n}  \label{4.7}
\end{equation}%
be an integral curve of the Hamiltonian vector field $X$, i.e.,
\begin{equation}
X(\alpha (t))=\overset{.}{\alpha },\,\,t\in \mathbf{R}.  \label{4.8}
\end{equation}%
In the local coordinates, it is obtained that
\begin{equation}
\alpha
(t)=(x_{i},x_{n+i},x_{2n+i},x_{3n+i},x_{4n+i},x_{5n+i},x_{6n+i},x_{7n+i})
\label{4.9}
\end{equation}%
and%
\begin{equation}
\begin{array}{c}
\overset{.}{\alpha }(t)=\frac{dx_{i}}{dt}\frac{\partial }{\partial x_{i}}+%
\frac{dx_{n+i}}{dt}\frac{\partial }{\partial x_{n+i}}+\frac{dx_{2n+i}}{dt}%
\frac{\partial }{\partial x_{2n+i}}+\frac{dx_{3n+i}}{dt}\frac{\partial }{%
\partial x_{3n+i}} \\
+\frac{dx_{4n+i}}{dt}\frac{\partial }{\partial x_{4n+i}}+\frac{dx_{5n+i}}{dt}%
\frac{\partial }{\partial x_{5n+i}}+\frac{dx_{6n+i}}{dt}\frac{\partial }{%
\partial x_{6n+i}}+\frac{dx_{7n+i}}{dt}\frac{\partial }{\partial x_{7n+i}}.%
\end{array}
\label{4.10}
\end{equation}%
Considering \textbf{Eq. }(\ref{4.8}), if equaled \textbf{Eq. }(\ref{4.6}) and%
\textbf{\ Eq. }(\ref{4.10}), it follows%
\begin{equation}
\begin{array}{c}
\frac{dx_{i}}{dt}=-\frac{\partial \mathbf{H}}{\partial x_{n+i}},\text{ }%
\frac{dx_{n+i}}{dt}=\frac{\partial \mathbf{H}}{\partial x_{i}},\text{ }\frac{%
dx_{2n+i}}{dt}=-\frac{\partial \mathbf{H}}{\partial x_{4n+i}},\text{ }\frac{%
dx_{3n+i}}{dt}=-\frac{\partial \mathbf{H}}{\partial x_{5n+i}}, \\
\frac{dx_{4n+i}}{dt}=\frac{\partial \mathbf{H}}{\partial x_{2n+i}},\text{ }%
\frac{dx_{5n+i}}{dt}=\frac{\partial \mathbf{H}}{\partial x_{3n+i}},\text{ }%
\frac{dx_{6n+i}}{dt}=-\frac{\partial \mathbf{H}}{\partial x_{7n+i}},\text{ }%
\frac{dx_{7n+i}}{dt}=\frac{\partial \mathbf{H}}{\partial x_{6n+i}}.%
\end{array}
\label{4.11}
\end{equation}%
Thus, the equations obtained in \textbf{Eq. }(\ref{4.11}) are seen to be
\textit{Hamiltonian equations} with respect to component $J_{1}^{\ast }$ of
almost Cliffordian structure $V^{\ast }$ on Cliffordian K\"{a}hler manifold $%
(\mathbf{R}^{8n},V),$ and then the triple $(\mathbf{R}^{8n},\Phi
_{J_{1}^{\ast }},X)$ is seen to be a \textit{Hamiltonian mechanical system }%
on Cliffordian K\"{a}hler manifold $(\mathbf{R}^{8n},V)$.

Secondly, let $(\mathbf{R}^{8n},V)$ be a Cliffordian K\"{a}hler manifold.
Suppose that an element of almost Cliffordian structure $V^{\ast }$, a
Liouville form and a 1-form on Cliffordian K\"{a}hler manifold $(\mathbf{R}%
^{8n},V)$ are denoted by $J_{2}^{\ast }$, $\lambda _{J_{2}^{\ast }}$ and $%
\omega _{J_{2}^{\ast }}$, respectively$.$

Putting
\begin{eqnarray*}
\omega _{J_{2}^{\ast }} &=&\frac{1}{2}%
(x_{i}dx_{i}+x_{n+i}dx_{n+i}+x_{2n+i}dx_{2n+i}+x_{3n+i}dx_{3n+i} \\
&&+x_{4n+i}dx_{4n+i}+x_{5n+i}dx_{5n+i}+x_{6n+i}dx_{6n+i}+x_{7n+i}dx_{7n+i})
\end{eqnarray*}%
we have
\begin{eqnarray*}
\lambda _{J_{2}^{\ast }} &=&J_{2}^{\ast }(\omega _{J_{2}^{\ast }})=\frac{1}{2%
}(x_{i}dx_{2n+i}-x_{n+i}dx_{4n+i}-x_{2n+i}dx_{i}+x_{3n+i}dx_{6n+i} \\
&&+x_{4n+i}dx_{n+i}-x_{5n+i}dx_{7n+i}-x_{6n+i}dx_{3n+i}+x_{7n+i}dx_{5n+i}).
\end{eqnarray*}

Assume that $X$ is a Hamiltonian vector field related to Hamiltonian energy $%
\mathbf{H}$ and given by \textbf{Eq. }(\ref{4.2}).

Considering
\begin{equation}
\Phi _{J_{2}^{\ast }}=-d\lambda _{J_{2}^{\ast }}=dx_{n+i}\wedge
dx_{4n+i}+dx_{2n+i}\wedge dx_{i}+dx_{5n+i}\wedge dx_{7n+i}+dx_{6n+i}\wedge
dx_{3n+i},  \label{4.12}
\end{equation}%
then we calculate%
\begin{equation}
\begin{array}{c}
i_{X}\Phi _{J_{2}^{\ast }}=\Phi _{J_{2}^{\ast
}}(X)=X^{n+i}dx_{4n+i}-X^{4n+i}dx_{n+i}+X^{2n+i}dx_{i}-X^{i}dx_{2n+i} \\
+X^{5n+i}dx_{7n+i}-X^{7n+i}dx_{5n+i}+X^{6n+i}dx_{3n+i}-X^{3n+i}dx_{6n+i}.%
\end{array}
\label{4.13}
\end{equation}%
According to \textbf{Eq.}(\ref{1.1}), if we equal \textbf{Eq. }(\ref{4.5})
and \textbf{Eq. }(\ref{4.13}), it follows%
\begin{equation}
\begin{array}{c}
X=-\frac{\partial \mathbf{H}}{\partial x_{2n+i}}\frac{\partial }{\partial
x_{i}}+\frac{\partial \mathbf{H}}{\partial x_{4n+i}}\frac{\partial }{%
\partial x_{n+i}}+\frac{\partial \mathbf{H}}{\partial x_{i}}\frac{\partial }{%
\partial x_{2n+i}}-\frac{\partial \mathbf{H}}{\partial x_{6n+i}}\frac{%
\partial }{\partial x_{3n+i}} \\
-\frac{\partial \mathbf{H}}{\partial x_{n+i}}\frac{\partial }{\partial
x_{4n+i}}+\frac{\partial \mathbf{H}}{\partial x_{7n+i}}\frac{\partial }{%
\partial x_{5n+i}}+\frac{\partial \mathbf{H}}{\partial x_{3n+i}}\frac{%
\partial }{\partial x_{6n+i}}-\frac{\partial \mathbf{H}}{\partial x_{5n+i}}%
\frac{\partial }{\partial x_{7n+i}}%
\end{array}
\label{4.14}
\end{equation}

Considering \textbf{Eq. }(\ref{4.8}), \textbf{Eq. }(\ref{4.10}) and\textbf{\
Eq. }(\ref{4.14}) are equal, we find equations%
\begin{equation}
\begin{array}{c}
\frac{dx_{i}}{dt}=-\frac{\partial \mathbf{H}}{\partial x_{2n+i}},\text{ }%
\frac{dx_{n+i}}{dt}=\frac{\partial \mathbf{H}}{\partial x_{4n+i}},\text{ }%
\frac{dx_{2n+i}}{dt}=\frac{\partial \mathbf{H}}{\partial x_{i}},\text{ }%
\frac{dx_{3n+i}}{dt}=-\frac{\partial \mathbf{H}}{\partial x_{6n+i}}, \\
\frac{dx_{4n+i}}{dt}=-\frac{\partial \mathbf{H}}{\partial x_{n+i}},\text{ }%
\frac{dx_{5n+i}}{dt}=\frac{\partial \mathbf{H}}{\partial x_{7n+i}},\text{ }%
\frac{dx_{6n+i}}{dt}=\frac{\partial \mathbf{H}}{\partial x_{3n+i}},\text{ }%
\frac{dx_{7n+i}}{dt}=-\frac{\partial \mathbf{H}}{\partial x_{5n+i}}.%
\end{array}
\label{4.15}
\end{equation}%
In the end, the equations obtained in \textbf{Eq. }(\ref{4.15}) are known to
be \textit{Hamiltonian equations} with respect to component $J_{2}^{\ast }$
of \ standard almost Cliffordian structure $V^{\ast }$ on the standard
Cliffordian K\"{a}hler manifold $(\mathbf{R}^{8n},V),$ and then the triple $(%
\mathbf{R}^{8n},\Phi _{J_{2}^{\ast }},X)$ is a \textit{Hamiltonian
mechanical system }on the standard Cliffordian K\"{a}hler manifold $(\mathbf{%
R}^{8n},V)$.

Thirdly, let $(\mathbf{R}^{8n},V)$ be a standard Cliffordian K\"{a}hler
manifold. By $J_{3}^{\ast }$ $\lambda _{J_{3}^{\ast }}$ and $\omega
_{J_{3}^{\ast }},$ we denote a component of almost Cliffordian structure $%
V^{\ast }$, a Liouville form and a 1-form on Cliffordian K\"{a}hler manifold
$(\mathbf{R}^{8n},V)$, respectively$.$

Let $\omega _{J_{3}^{\ast }}$ be given by

\begin{eqnarray*}
\omega _{J_{3}^{\ast }} &=&\frac{1}{2}%
(x_{i}dx_{i}+x_{n+i}dx_{n+i}+x_{2n+i}dx_{2n+i}+x_{3n+i}dx_{3n+i} \\
&&+x_{4n+i}dx_{4n+i}+x_{5n+i}dx_{5n+i}+x_{6n+i}dx_{6n+i}+x_{7n+i}dx_{7n+i})
\end{eqnarray*}%
Then it holds

\begin{eqnarray*}
\lambda _{J_{2}^{\ast }} &=&J_{2}^{\ast }(\omega _{J_{2}^{\ast }})=\frac{1}{2%
}(x_{i}dx_{3n+i}-x_{n+i}dx_{5n+i}-x_{2n+i}dx_{6n+i}-x_{3n+i}dx_{i} \\
&&+x_{4n+i}dx_{7n+i}+x_{5n+i}dx_{n+i}+x_{6n+i}dx_{2n+i}-x_{7n+i}dx_{4n+i}).
\end{eqnarray*}%
It is well-known that if $\Phi _{J_{3}^{\ast }}$ is a closed K\"{a}hler form
on the standard Cliffordian K\"{a}hler manifold $(\mathbf{R}^{8n},V),$ then $%
\Phi _{J_{3}^{\ast }}$ is also a symplectic structure on Cliffordian K\"{a}%
hler manifold $(\mathbf{R}^{8n},V)$.

Consider $X$ . It is Hamiltonian vector field connected with Hamiltonian
energy $\mathbf{H}$ and given by \textbf{Eq. }(\ref{4.2}).

Taking into
\begin{equation}
\Phi _{J_{3}^{\ast }}=-d\lambda _{J_{3}^{\ast }}=dx_{3n+i}\wedge
dx_{i}+dx_{n+i}\wedge dx_{5n+i}+dx_{2n+i}\wedge dx_{6n+i}+dx_{7n+i}\wedge
dx_{4n+i},  \label{4.17}
\end{equation}%
we find%
\begin{equation}
\begin{array}{c}
i_{X}\Phi _{J_{3}^{\ast }}=\Phi _{J_{3}^{\ast
}}(X)=X^{3n+i}dx_{i}-X^{i}dx_{3n+i}+X^{n+i}dx_{5n+i}-X^{5n+i}dx_{n+i} \\
+X^{2n+i}dx_{6n+i}-X^{6n+i}dx_{2n+i}+X^{7n+i}dx_{4n+i}-X^{4n+i}dx_{7n+i}.%
\end{array}
\label{4.18}
\end{equation}%
According to \textbf{Eq.}(\ref{1.1}), \textbf{Eq. }(\ref{4.5}) and \textbf{%
Eq. }(\ref{4.18}) are equaled, we obtain a Hamiltonian vector field given by%
\begin{equation}
\begin{array}{c}
X=-\frac{\partial \mathbf{H}}{\partial x_{3n+i}}\frac{\partial }{\partial
x_{i}}+\frac{\partial \mathbf{H}}{\partial x_{5n+i}}\frac{\partial }{%
\partial x_{n+i}}+\frac{\partial \mathbf{H}}{\partial x_{6n+i}}\frac{%
\partial }{\partial x_{2n+i}}+\frac{\partial \mathbf{H}}{\partial x_{i}}%
\frac{\partial }{\partial x_{3n+i}} \\
-\frac{\partial \mathbf{H}}{\partial x_{7n+i}}\frac{\partial }{\partial
x_{4n+i}}-\frac{\partial \mathbf{H}}{\partial x_{n+i}}\frac{\partial }{%
\partial x_{5n+i}}-\frac{\partial \mathbf{H}}{\partial x_{2n+i}}\frac{%
\partial }{\partial x_{6n+i}}+\frac{\partial \mathbf{H}}{\partial x_{4n+i}}%
\frac{\partial }{\partial x_{7n+i}}.%
\end{array}
\label{4.19}
\end{equation}

Taking into \textbf{Eq. }(\ref{4.8}), we equal \textbf{Eq. }(\ref{4.10}) and%
\textbf{\ Eq. }(\ref{4.19}), it yields%
\begin{equation}
\begin{array}{c}
\frac{dx_{i}}{dt}=-\frac{\partial \mathbf{H}}{\partial x_{3n+i}},\text{ }%
\frac{dx_{n+i}}{dt}=\frac{\partial \mathbf{H}}{\partial x_{5n+i}},\text{ }%
\frac{dx_{2n+i}}{dt}=\frac{\partial \mathbf{H}}{\partial x_{6n+i}},\text{ }%
\frac{dx_{3n+i}}{dt}=\frac{\partial \mathbf{H}}{\partial x_{i}}, \\
\frac{dx_{4n+i}}{dt}=-\frac{\partial \mathbf{H}}{\partial x_{7n+i}},\text{ }%
\frac{dx_{5n+i}}{dt}=-\frac{\partial \mathbf{H}}{\partial x_{n+i}},\text{ }%
\frac{dx_{6n+i}}{dt}=-\frac{\partial \mathbf{H}}{\partial x_{2n+i}},\text{ }%
\frac{dx_{7n+i}}{dt}=\frac{\partial \mathbf{H}}{\partial x_{4n+i}}.%
\end{array}
\label{4.20}
\end{equation}%
Finally, the equations obtained in \textbf{Eq. }(\ref{4.20}) are obtained to
be \textit{Hamiltonian equations} with respect to component $J_{3}^{\ast }$
of almost Cliffordian structure $V^{\ast }$ on the standard Cliffordian K%
\"{a}hler manifold $(\mathbf{R}^{8n},V),$ and then the triple $(\mathbf{R}%
^{8n},\Phi _{J_{3}^{\ast }},X)$ is a \textit{Hamiltonian mechanical system }%
on the standard Cliffordian K\"{a}hler manifold $(\mathbf{R}^{8n},V)$.

\section{Conclusion}

Formalism of Hamiltonian mechanics has intrinsically been described with
taking into account the basis $\{J_{1}^{\ast },J_{2}^{\ast },J_{3}^{\ast }\}$
of almost Cliffordian structure $V^{\ast }$ on the standard Cliffordian K%
\"{a}hler manifold $(\mathbf{R}^{8n},V)$.

Hamiltonian models arise to be a very important tool since they present a
simple method to describe the model for mechanical systems. In solving
problems in classical mechanics, the rotational mechanical system will then
be easily usable model.

Since physical phenomena, as well-known, do not take place all over the
space, a new model for dynamic systems on subspaces is needed. Therefore,
equations (\ref{4.11}), (\ref{4.15}) and (\ref{4.20}) are only considered to
be a first step to realize how Cliffordian geometry has been used in solving
problems in different physical area.

For further research, the Hamiltonian vector fields derived here are
suggested to deal with problems in electrical, magnetical and gravitational
fields of quantum and classical mechanics of physics.

\end{document}